\documentstyle[emulateapj,flushrt,tighten]{article} %preprint
\input{psfig.sty}
\singlespace
\newcommand{\msun}{\,{\rm M_\odot}}

\newcommand{\etal}{{et al.\ }}

\newcommand{\ba}{\begin{eqnarray}}
\newcommand{\ea}{\end{eqnarray}}
\def\spose#1{\hbox to 0pt{#1\hss}}
\newcommand{\lta}{\mathrel{\spose{\lower 3pt\hbox{$\mathchar"218$}}
      \raise 2.0pt\hbox{$\mathchar"13C$}}}
\newcommand{\gta}{\mathrel{\spose{\lower 3pt\hbox{$\mathchar"218$}}
      \raise 2.0pt\hbox{$\mathchar"13E$}}}
\newcommand{\Lya}{Ly$\alpha\ $}
\newcommand{\lya}{Ly$\alpha\ $}
\def\lyac{Ly$\alpha$--c~}

\def\HI{\hbox{H~$\scriptstyle\rm I\ $}}
\def\HII{\hbox{H~$\scriptstyle\rm II\ $}}

\def\nH{{\rm H}}

\def\be{\begin{equation}}
\def\ee{\end{equation}}

\newcommand{\Junits}{{\rm\,ergs\,cm^{-2}\,s^{-1}\,Hz^{-1}\,sr^{-1}}}
\def\gtsima{$\; \buildrel > \over \sim \;$}
\def\ltsima{$\; \buildrel < \over \sim \;$}
\def\prosima{$\; \buildrel \propto \over \sim \;$}
\def\gsim{\lower.5ex\hbox{\gtsima}}
\def\lsim{\lower.5ex\hbox{\ltsima}}
\def\simgt{\lower.5ex\hbox{\gtsima}}
\def\simlt{\lower.5ex\hbox{\ltsima}}
\def\simpr{\lower.5ex\hbox{\prosima}}

\def\ie{{\frenchspacing i.e. }}

\lefthead{CIARDI \& MADAU}
\righthead{PROBING BEYOND THE REIONIZATION EPOCH}
\submitted{submitted to the ApJ}
\makeatletter

\newenvironment{figurehere}
  {\def\@captype{figure}}
  {}
\makeatother

\begin{document}

\title{Probing beyond the epoch of hydrogen reionization with 21 centimeter
radiation}

\author{Benedetta Ciardi\altaffilmark{1} \& Piero Madau\altaffilmark{2,3}}

\altaffiltext{1}{Max-Planck-Institut f\"ur Astrophysik, 
Karl-Schwarzschild-Stra\ss e 1, 85748 Garching, Germany.}
\altaffiltext{2}{Department of Astronomy \& Astrophysics, University of 
California, Santa Cruz, CA 95064.}
\altaffiltext{3}{Osservatorio Astrofisico di Arcetri, Largo E. Fermi 5,
50125 Firenze, Italy.}

\received{---------------}
\accepted{---------------}
 
\begin{abstract}

We use numerical simulations of hydrogen reionization by stellar sources 
in the context of $\Lambda$CDM cosmogonies to investigate the 21 $(1+z)$ cm
radio signal expected from the diffuse intergalactic medium (IGM)
prior to the epoch of reionization breakthrough at redshift $z_{\rm ion}$. 
Two reionization scenarios are analyzed in
details: an `early reionization' case with $z_{\rm ion}\approx 13$,
consistent with the recent discovery by the {\it WMAP} satellite of a
large optical depth to Thomson scattering, and a `late reionization'
case with $z_{\rm ion}\approx 8$. It is a generic prediction
of these models that the background of \Lya photons produced by the early 
generation of stars which ultimately ionize the universe  
will be sufficiently intense to make intergalactic neutral hydrogen visible 
against the cosmic microwave background during the `grey age', \ie $z_{\rm ion}
\lta z \lta 20$. 
Depending on the redshift of reionization breakthrough, broad-beam observations
at frequencies $\lta 150$ MHz (below 100 MHz for the `early reionization'
scenario) with the next generation of radio telescopes 
should reveal angular fluctuations in the sky brightness temperature in the 
range 5--20 mK (1$\sigma$) on scales below 5 arcmin. 
\end{abstract}


\section{Introduction}

Despite much recent theoretical and observational progress in our 
understanding of the formation of early cosmic structures and the high 
redshift universe, many fundamental questions remain, at best, only 
partially answered. We know that at least some galaxies and quasars were
already shining when the universe was less than $10^9$ yr old. But 
when did the first luminous objects form, and what was their impact on the
surrounding intergalactic gas? While the excess \HI absorption measured in 
the spectra of $z\sim 6$ quasars in the Sloan
Digital Sky Survey has been interpreted as the signature of the trailing 
edge of the cosmic reionization epoch (Becker \etal 2001; Fan \etal 2002), 
the recent analysis of the first year data from the {\it Wilkinson Microwave 
Anisotropy Probe} ({\it WMAP}) satellite infers a mean optical depth to 
Thomson scattering $\tau_e \sim 0.17$, suggesting that the universe was 
reionized at higher redshift (Kogut \etal 2003; Spergel \etal 2003). This
is an indication of significant star-formation activity at very early times. 
In currently popular cosmological scenarios 
it is an early generation of subgalactic stellar systems -- aided perhaps by a 
seed population of accreting black holes in their nuclei -- that may have 
generated the ultraviolet radiation that ended the cosmic `dark ages' and 
reionized most of the hydrogen in the universe around redshift 8--15. The 
detailed thermal
history of the baryons during and soon after these crucial formative 
stages depends on the 
power-spectrum of density fluctuations on small scales and on a complex 
network of poorly understood `feedback' mechanisms, and remains one of 
the missing link in galaxy formation and evolution studies. 

An alternative way to probe the end of the cosmic `dark ages' and discriminate
between different reionization histories is through 21 cm 
tomography. It has also long been known (e.g.
Field 1959; Sunyaev \& Zel'dovich 1975; Hogan \& Rees 1979; 
Scott \& Rees 1990; Subramanian 
\& Padmanabhan 1993) that neutral hydrogen in the intergalactic medium 
(IGM) and gravitationally collapsed systems may be directly 
detectable in emission or absorption against the cosmic microwave background
radiation (CMB) at the frequency corresponding to the redshifted 
21 cm line (associated with the spin-flip transition from the 
triplet to the singlet state). Madau, Meiksin, \& Rees (1997, hereafter 
MMR) first showed that 21 cm tomography could provide a direct probe of the 
era of cosmological reionization and reheating. In general, 
21 cm spectral features will display angular structure as well as structure 
in redshift space due to inhomogeneities in the gas density field, 
hydrogen ionized fraction, and spin temperature. Several different 
signatures have been 
investigated in the recent literature: (a) the fluctuations in the 
21 cm line
emission induced by the `cosmic web' that develops 
at early times in cold dark matter (CDM)-dominated cosmologies 
(Tozzi \etal 2000) and by 
`minihalos' with virial temperatures below $10^4\,$K 
(Iliev \etal 2002,2003); (b) the global feature (`reionization step') in the
continuum spectrum of the radio sky that may mark the abrupt overlapping
phase of individual intergalactic \HII regions (Shaver \etal 1999); 
(c) and the 21 cm narrow lines generated in absorption against very 
high redshift radio sources 
by the neutral IGM (Carilli, Gnedin, \& Owen 2002) and by intervening 
minihalos  and protogalactic disks (Furlanetto \& Loeb 2002).

In this paper, we use numerical simulations of hydrogen reionization 
by stellar sources in the context of $\Lambda$CDM cosmogonies 
to estimate the radio signal expected from the diffuse, low density
IGM during the `gray age' (Carilli \etal 2002). This is the era between 
the formation of the first luminous objects and the epoch of complete 
reionization,
when the IGM is a mixture of neutral, partially ionized, and fully ionized 
structures. The important point is that the early generation of stars 
that ultimately
ionizes the universe will also generate a background radiation field of 
\lya photons arising from their collective redshifted UV continua.
It is the presence of a sufficient flux of \Lya photons that renders the 
neutral IGM `visible' against the CMB by mixing the hyperfine levels
(`\Lya pumping'). Radio maps will show a 
patchwork (both in angle and in frequency) of emission signals from \HI 
zones modulated by \HII regions where no signal is detectable against the 
CMB. 

The remainder of this paper is organized as follows. In \S\,2 we summarize 
the physics of 21 cm excitation. In \S\,3 we show that, since a `typical' 
early stellar population will radiate a large number 
of UV continuum photons with energies between 10.2 and 13.6 eV (hereafter \lyac) to
which the IGM is transparent, there will be an extended period of time 
prior to full reionization 
where fluctuations in the gas density and hydrogen ionized fraction may 
be detected at $21\,(1+z)\,$cm. Numerical simulations of hydrogen reionization
by Population III stars are presented in \S\,4. Simulated radio maps are 
generated 
and the detectability of fluctuations in the sky brightness discussed in \S\,5.
Finally, in \S\,6 we give our conclusions.     

\section{Lyman-alpha pumping}

A quick summary of the physics of 21 cm radiation will illustrate 
the basic ideas behind this work (see MMR for more details).
The emission or absorption of 21 cm photons from a neutral IGM is
governed by the hydrogen spin temperature, $T_S$, defined by 
\be
n_1/n_0=3\exp(-T_*/T_S), \label{eqts}
\ee
where $n_0$ and $n_1$ are the number densities of atoms in the singlet and 
triplet $n=1$ hyperfine levels
and $T_*=0.068\,$K is the excitation temperature of the transition. 
In the presence of only the CMB radiation with
$T_{\rm CMB}=2.73\,(1+z)\,$K, the spin states will reach thermal equilibrium
with the CMB on a timescale of $T_*/(T_{\rm CMB}A_{10})\approx 3\times10^5\,
(1+z)^{-1}\,$yr ($A_{10}=2.85\times 10^{-15}\,$s$^{-1}$
is the spontaneous decay rate of the hyperfine transition of atomic hydrogen),
and intergalactic \HI will produce neither an absorption nor an emission
signature. Thus, a mechanism is required that decouples $T_S$ and $T_{\rm CMB}$,
e.g. by coupling the spin temperature instead to the kinetic temperature
$T_K$ of the gas itself. Two mechanisms are available, spin-exchange 
collisions between hydrogen atoms (Purcell \& Field 1956) and scattering 
by \Lya photons (Wouthuysen 1952; Field 1958; Field 1959). The first process 
proceeds at a rate that is too small for realistic IGM
densities at the redshifts of interest, although collisions will be
important in dense regions with gas overdensities $\delta\gta 20[(1+z)/10]^{-2}$,
like virialized minihalos (Iliev \etal 2002,2003; Furlanetto \& Loeb 2002). 
It is \Lya pumping that dominates instead in the diffuse IGM, 
by mixing the hyperfine levels of neutral hydrogen in its
ground state via intermediate transitions to the $2p$ state, the
Wouthuysen-Field process. 
In this case,
\begin{equation}
P_\alpha={4\pi J_\alpha \sigma_\alpha\over h_P\nu_\alpha},
\end{equation}
is the rate at which \Lya photons of energy $h_P\nu_\alpha$ in an isotropic
background of intensity $J_\alpha$ are scattered by an H atom in the gas, and
\begin{equation}
P_{\rm th}\equiv \frac {27 A_{10} T_{\rm CMB}}{4 T_*}=
7.6 \times 10^{-13}\,(1+z)~{\rm s}^{-1},
\end{equation}
is the critical (`thermalization') scattering rate which, if much exceeded, 
would drive $T_S$ away from $T_{\rm CMB}$ and towards $T_K$ (MMR). 
In the above equations $\sigma_\alpha=\int \sigma_\nu d\nu=\pi e^2 
f_\alpha/m_e c=0.011\,$cm$^{2}\,$Hz
is the integrated Ly$\alpha$ scattering cross section.
The condition $P_\alpha>P_{\rm th}$ can also be rewritten as
\begin{equation}
J_\alpha>J_{\rm th}=9\times 10^{-23} (1+z)\;\Junits.
\label{jstrong}
\end{equation}
At $z=8$ this thermalization rate corresponds to $P_{\rm th}\nu_\alpha
/(cn_\nH \sigma_\alpha)\lta 1/2$ \Lya photons per hydrogen atom, 
where $n_\nH$ is the intergalactic mean hydrogen density. 

\section{UV spectra of early stellar populations}

At the epochs of interest, \Lya pumping will efficiently decouple $T_S$ 
from $T_{\rm CMB}$
if $J_\alpha\gta 10^{-21}\,$ergs cm$^{-2}$ s$^{-1}$ Hz$^{-1}$ 
sr$^{-1}$. The issue is then whether this critical value of $J_\alpha$ will be
reached well before the epoch of complete reionization, when large regions 
of the IGM are still neutral. 
The process of reionization begins as individual sources drive isolated 
\HII regions in the surrounding IGM and ends when these ionized bubbles
overlap and fill the intergalactic volume.
The volume filling factor of \HII regions in a clumpy medium reaches unity when about 
$1+t/\bar t_{\rm rec}\,$ ionizing photons have been emitted per hydrogen atom, where
$\bar t_{\rm rec}$ is the volume--averaged hydrogen recombination timescale (Madau, 
Haardt, \& Rees 1999). Based on the density field in large scale cosmological 
simulations, only a few Lyman-continuum (Lyc) photons per hydrogen atom may be needed 
(Gnedin 2000; Miralda-Escud\'e, Haehnelt, \& Rees 2000) to keep the gas in mildly 
overdense regions and intergalactic filaments ionized against radiative recombinations 
at $z\sim 7$; many photons, however, will 
not escape local absorption, i.e. only a small fraction, $f_{\rm esc}$, of the Lyc 
radiation emitted
by massive stars is expected to escape the dense sites of star formation into galaxy 
halos and the intergalactic space. Moreover, at the earliest epochs of 
structure formation 
in CDM cosmologies the smallest nonlinear objects are the numerous halos that 
condense with masses just above the cosmological Jeans mass. Such `minihalos'
are not yet fully resolved nor is the process of their photoevaporation 
captured in large-scale three-dimensional cosmological simulations. This 
sub-grid structure
makes the treatment of IGM clumping in all presently available simulations only
approximate (Haiman, Abel, \& Madau 2001). The error in small-scale clumping
(that leads to underestimating the total recombination rate) can be partly
compensated by the assumption of a small escape fraction.

The early generation of massive stars likely responsible for reionization will also 
generate a background radiation field of photons with energies between 10.2--13.6 eV
to which the IGM is transparent. As each of these \lyac photons gets redshifted,
it will ultimately reach the \lya transition energy of 10.2 eV, scatter
resonantly off neutral hydrogen, and mix the hyperfine levels. 
The observability of the pre-reionization IGM depends therefore on the 
UV spectrum of the firsts stars, i.e. on the number of \lyac photons 
emitted per H-ionizing photon, and on the escape fraction of Lyc photons into the IGM.
Figure ~\ref{fig1} shows the number of continuum photons per octave near the \lya 
frequency emitted by a stellar population, $N_\alpha$. The UV spectral energy 
distribution is characterized by a strong Lyc break the size 
of which depends on age, initial mass function (IMF), metallicity, and star formation
history. We have therefore normalized $N_\alpha$ to the total number of 
hydrogen-ionizing photons emitted, $N_{\rm ion}$. The values shown have 
been computed assuming a simple 
stellar population of metal-free `Population III' stars: the
simulations described in the next section use synthetic, time-dependent 
spectra of zero-metallicity stars to model the emission 
properties of stellar sources. For comparison, the figure also depicts the 
number of \lya photons emitted by low metallicity stars ($Z=Z_\odot/50$).
Note how the hotter temperatures of Population III stars produce a smaller Lyc 
break hence a smaller $N_\alpha/N_{\rm ion}$ ratio.
Also, after integrating over stellar age, a starburst is characterized by a 
ratio $N_\alpha/N_{\rm ion}$ that is a few times higher than in the constant 
star formation rate case (this is because of the shorter lifetime of 
the massive stars that produce radiation above 1 ryd). 
It is possible that the starburst mode may be more relevant for the
galaxies responsible for reionization, as stellar winds and supernovae can
easily expel the gas out of the shallow potential wells of low mass halos, after 
the first burst of star formation occurs (Madau, Ferrara, \& Rees 2001). 

\begin{figurehere}
\vspace{-0.3cm}
\centerline{
\psfig{figure=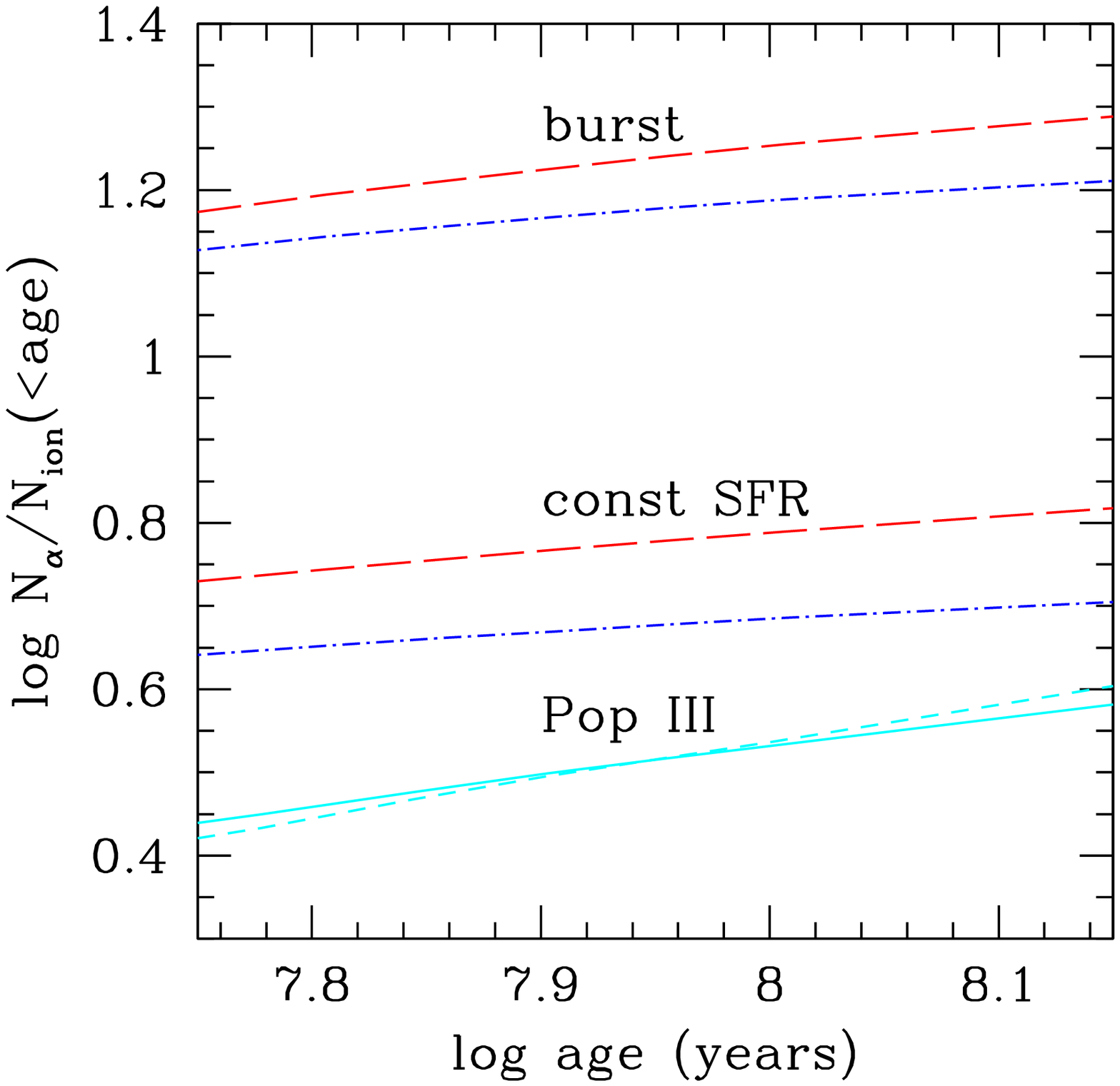,width=2.8in}}
\caption{\footnotesize
Number of continuum photons per octave in frequency just shortward of Ly$\alpha$,
$N_\alpha$, emitted as a 
function of age by a stellar population with a Salpeter
IMF ({\it dot-dashed lines)} and a Scalo ({\it long-dashed lines}) IMF. Upper set of
curves assumes an instantaneous burst of star formation, lower set a constant
star formation rate. We have normalized $N_\alpha$ to the number of hydrogen-ionizing 
photons per octave emitted just shortward of the Lyman limit, $N_{\rm 
ion}$. The population synthesis values have been computed for low
metallicities ($Z=0.02\,Z_\odot$), and are based on an update of Bruzual \&
Charlot's (1993) libraries. {\it Solid line}: same for zero-metallicity 
Population III stars with a Salpeter IMF. {\it Short-dashed line}: same for 
zero-metallicity Population III stars with a Larson IMF (see Ciardi \etal 
2001 and references therein).
}
\label{fig1}
\vspace{0.6cm}
\end{figurehere}

As mentioned above, only an amount $N_{\rm phot}=f_{\rm esc}\,N_{\rm ion}$ 
of Lyc photons will 
actually be able to escape from the galaxies and break into the IGM. A smaller escape 
fraction will generate a stronger \Lya diffuse radiation background for a given 
ionizing metagalactic flux. Taking a fiducial ratio $N_\alpha/N_{\rm ion}\gta 
4$ and a few H-ionizing photons per atom to keep the IGM fully photoionized, 
$N_{\rm phot}/N_\nH\approx 5$ (see Fig. 7 in Ciardi, Stoehr, \& White 
2003, hereafter CSW), these simple estimates imply 
\be 
{N_\alpha\over N_\nH}\gta 4 {N_{\rm phot}\over N_\nH}\,f_{\rm esc}^{-1}\gta 
200\,(f_{\rm esc}/0.1)^{-1}
\ee
\lya `continuum' photons emitted per hydrogen atom close to the 
reionization epoch. This is more than two orders of magnitude larger 
than the thermalization rate for efficient level mixing by \Lya pumping. It 
appears then to be a generic prediction of cosmological
reionization scenarios by stellar sources that {\it the \lya background 
from the early generation of stars which ultimately ionize the 
universe should be sufficiently intense to make intergalactic diffuse 
neutral hydrogen visible against the CMB well before the epoch of full 
reionization.} 
As shown below, detailed simulations of IGM reionization that include 
cosmological radiative transfer effects will confirm this basic result.

\section{Numerical Simulations of IGM Reionization}

In this and the following section we use the numerical simulations
of hydrogen reionization by stellar sources described in CSW
and Ciardi, Ferrara, \& White (2003, hereafter CFW) to investigate 
the radio signal expected from the diffuse IGM prior to reionization.
We give here a brief summary of the main features of the
simulations, relevant to the present study.
In the above papers, the reionization process is studied through
a combination of high resolution N-body 
simulations (to describe the distribution of dark matter and diffuse gas,
Springel, Yoshida, \& White 2001; Yoshida, Sheth, \& 
Diaferio 2001), a semi-analytic
model of galaxy formation (to track gas cooling, star formation, and feedback
from supernovae, Kauffmann et al. 1999; Springel et al. 2001) and the
Monte Carlo radiative transfer code {\tt CRASH} (to follow the propagation 
of ionizing photons into the IGM, Ciardi et al. 2001; Maselli, Ferrara, 
\& Ciardi 2003). The simulations, spanning the redshift range $z=6-20$, 
are performed for a $\Lambda$CDM cosmology with 
($\Omega_m, \Omega_\Lambda, \sigma_8, h, \Omega_b, n)=(0.3, 0.7, 0.9, 
0.7, 0.04, 1)$. 
The simulation box has a comoving length of $L=20 h^{-1}$~Mpc and a particle
mass of $M_p=1.7 \times 10^8 h^{-1}$~M$_\odot$. This choice of the
parameters allows to simulate a region of the universe with `mean'  
properties, thus avoiding biases due to cosmic variance on small scales and, 
at the same time, to resolve the objects that produce the bulk of the
ionizing radiation (Ciardi 2002).
Each output of the simulation provides the gas distribution inside the
box together with a galaxy catalog containing, among other quantities, 
their positions, stellar masses, and star formation rates. As discussed
in \S~5, two sets of simulations will be used, with different choices
for the galaxy emission properties.  
At each output, the propagation into the IGM of the ionizing photons
emitted by the simulated galaxy population is followed with the radiative
transfer code {\tt CRASH}. The code has been tested to ensure a good 
convergence in the ionization fraction for a wide range of densities.
The calculation is performed over a $N=128^3$ grid.
The gas (pure hydrogen) distribution is assumed to track the dark matter.
While radiation above 13.6 eV from each source is fully absorbed at the 
ionization front surrounding it, the \lyac radiation is free to propagate and 
fill the space in between the \HII regions. Since numerous sources are contributing 
to the \lya background throughout the universe, its flux is nearly isotropic in 
the cosmic frame of reference, and is given by 
\begin{equation}
J_\alpha(z)=\frac {c} {4 \pi} \int_z^{z_{max}} \epsilon(\nu',z') \frac
{(1+z)^3} {(1+z')^3} \left\vert \frac{dt} {dz'} \right\vert dz',
\label{jalpha}
\end{equation}
where the proper emissivity $\epsilon(\nu',z')$ can be written as 
\begin{equation}
\epsilon(\nu',z')={(1+z')^3\over L^3}M_*(z')S(\nu').
\label{emiss}
\end{equation}
Here, $\nu'\equiv \nu_\alpha (1+z')/(1+z)$, $M_*(z')$ is the total stellar mass
in the simulation box at redshift $z'$, and $S(\nu')$ is the time-dependent 
spectral energy 
distribution at frequency $\nu'$. Since the UV photon production is rapidly 
decreasing 
with stellar age, $M_*(z')$ is the mass of newly (i.e. in the timescale between
two outputs of the simulation) formed stars, i.e. we neglect the contribution 
from stars formed in previous outputs. 
In equation (\ref{jalpha}) we have assumed that Ly$\alpha$ photons are not
attenuated by the intervening IGM, and that the IGM is opaque to ionizing photons prior
to reionization, $z>z_{\rm ion}$. Thus, the integration over redshift is carried out only
up to $1+z_{\rm max}=4 (1+z)/3$. 
%For redshift $z<z_{\rm ion}$ also photons above the Lyman limit 
% contribute to the background, although they are attenuated by galactic
%absorption. 
In Figure \ref{fig2} the resulting \Lya background intensity is depicted
as a function 
of redshift, together with the volume-averaged ionization fraction $x_v$,
for the two simulation runs.
The figure clearly shows that the expected diffuse flux of \lya photons 
should be sufficiently intense to decouple the spin temperature from the 
CMB during the `grey age' from redshift $\sim 20$ to complete reionization.

\begin{figurehere}
\vspace{-0.3cm}
\centerline{
\psfig{figure=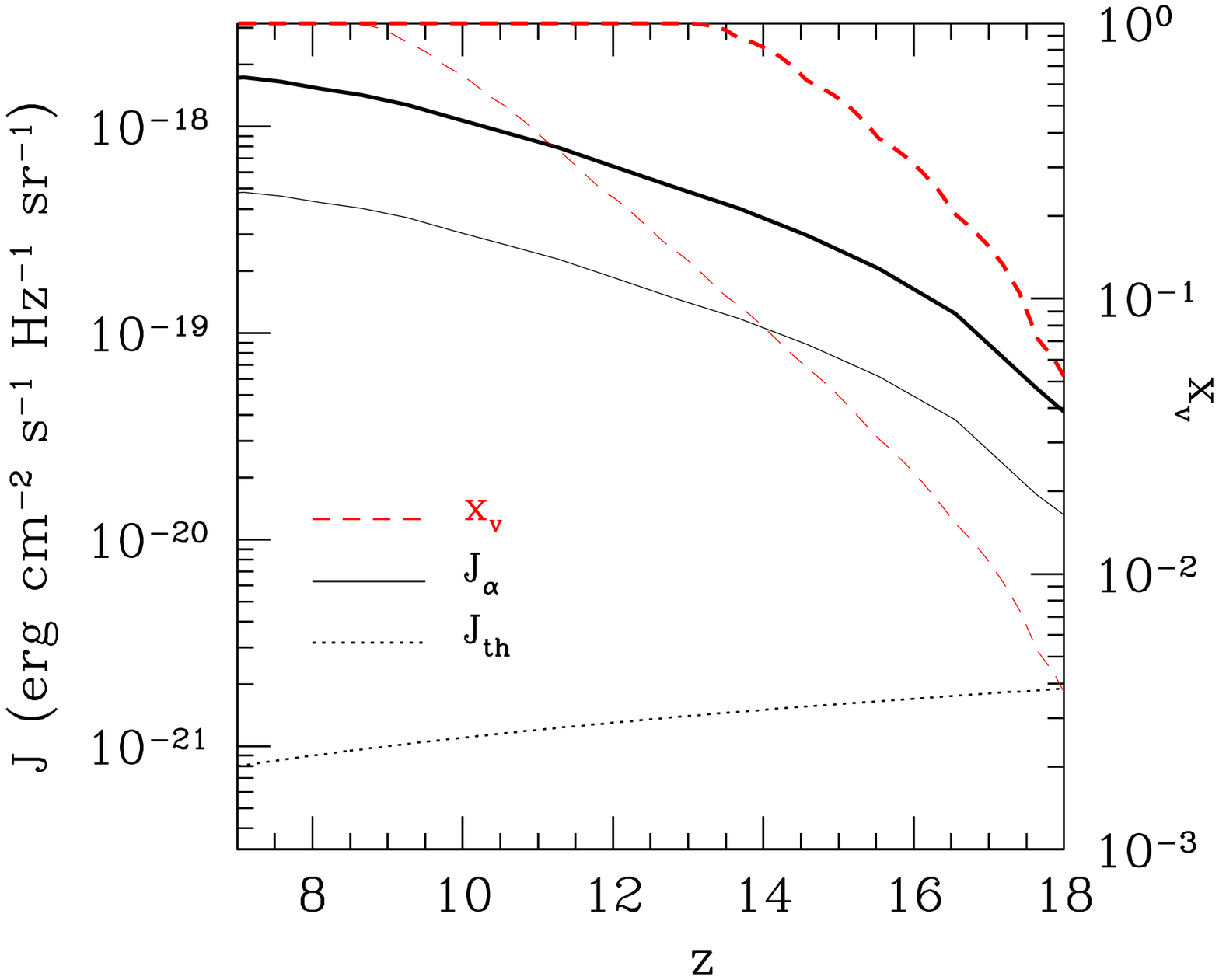,width=2.8in}}
\caption{\footnotesize {\it Solid lines}: redshift evolution of the `continuum' 
\lya background, $J_\alpha$, predicted by our simulations of cosmological 
reionization by stellar sources. {\it Dashed lines:} volume-averaged ionization 
fraction, $x_v$. For each set of curves, the thick (thin) line refers to the 
`early reionization' (`late reionization') scenario.
{\it Dotted line}: Thermalization rate, $J_{\rm th}$, needed to 
efficiently decouple the spin temperature of neutral hydrogen from the CMB 
via \Lya pumping (see text for details). 
}
\label{fig2}
\vspace{1.6cm}
\end{figurehere}

\section{Radio emission from the diffuse IGM}

A beam of 21 cm radiation passing through a neutral hydrogen patch causes absorption
and induces emission. In the radio range the exponential in equation (\ref{eqts})
is close to unity and stimulated emission cancels most of the 
absorption. The 21cm absorption coefficient depends then on the spin 
temperature. 
If the patch is located at redshift $z$, has (mean) spin temperature $T_S$,
cosmic overdensity $\delta$, neutral hydrogen fraction $(1-x)$, 
angular diameter on the sky larger than the beamwidth of the radio telescope,
and a radial velocity width broader than the bandwidth, its optical 
depth at $21 (1+z)\,$cm is given by (MMR)
\be
\tau=\frac{3}{32\pi}\lambda_{10}^3A_{10}\frac{T_*}{T_S}(1+\delta)(1-x)\,
\frac{n_\nH}{H}\simeq 0.006\,{\cal T}\,{T_{\rm CMB}\over T_S},
\label{tau}
\ee
where we have included in the function ${\cal T}$ all dependences on 
the density and ionization state of the patch, and on cosmological 
parameters,  
\be
{\cal T}\equiv {1\over h}(1+\delta) (1-x)\left({\Omega_bh^2\over 0.02}\right) 
\left[\left({1+z\over 10}\right)\left({0.3\over \Omega_m}\right)\right]
^{1/2}.
\label{taucosm}
\ee
Here the Hubble constant is $H_0=100\,h\,$ km s$^{-1}$ Mpc$^{-1}$ and we have 
used the approximation $H(z)\approx \sqrt{\Omega_m}H_0(1+z)^{3/2}$. 

\begin{figure*}
\centerline{
\psfig{figure=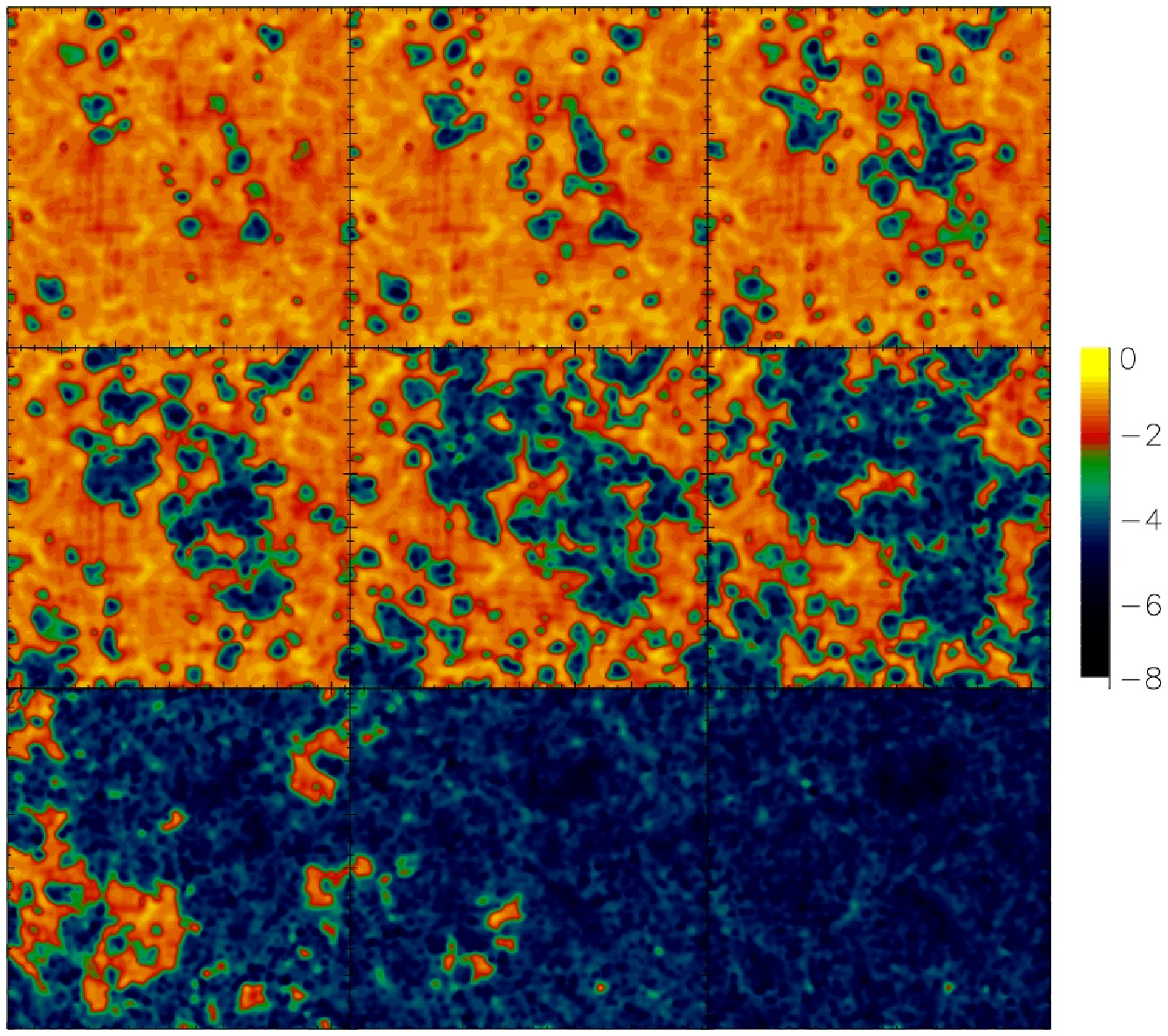,width=12.2cm}}
\vspace{0.8cm}
\caption{\footnotesize Slices through the simulation box in the `late 
reionization' model. The nine panels 
show the differential antenna temperature $\log\,(\delta T_b/$K) at redshifts
(from top to bottom and left to right) $z=$~13.5, 12.8, 12.0, 11.3, 
10.6, 9.9, 9.3, 8.7, and 8.1. The maps refer to the emission produced by 
a slice of gas of comoving thickness $L/N^{1/3}=220\,$kpc.
}
\label{fig3}
%\vspace{0.6cm}
\end{figure*}

Because of the
weakness of the magnetic hyperfine transition, $\tau$ will typically be 
much less than unity. In the rest frame of the patch, the radiative 
transfer
equation yields for the brightness temperature through the translucent IGM 
$T_b=T_{\rm CMB} e^{-\tau}+T_S(1-e^{-\tau})$. The flux observed at Earth can
be expressed by the differential antenna temperature between this patch and
the CMB, $\delta T_b=(T_b-T_{\rm CMB})(1+z)^{-1}$ or
\be
\delta T_b\simeq {T_S-T_{\rm CMB}\over 1+z}\,\tau
\simeq 0.016\,{\rm K}~{\cal T}~\left(1-{T_{\rm CMB}\over T_S}\right).
\label{eq:dT}
\ee
If $P_\alpha>P_{\rm th}$ and the IGM has been preheated by primordial 
sources of radiation, $T_S$ will be much larger than $T_{\rm CMB}$, and the 
universe will be observable in 21 cm emission at a level that is independent 
of the exact value of $T_S$. If preheating is negligible, the adiabatic 
expansion of the universe will lower the kinetic (hence spin) temperature 
of the gas 
well below that of the CMB, and the IGM will be detectable in absorption 
against the CMB.
The energetic demand for heating the IGM above the CMB temperature is meager,
only $\sim 0.004\,$ eV per particle at $z\sim 10$. Consequently, even
relatively inefficient heating mechanisms may be important warming sources as
structure develops before the universe was actually reionized, such as 
\Lya heating (MMR) and X-ray heating (MMR; Venkatesan, Giroux, \& Shull 2001)
in low density regions, and shock heating of the gas in high density regions 
(Carilli \etal 2002). In the cosmological simulations of Carilli \etal (2002),
the IGM is reionized `late' but the volume-averaged kinetic and spin 
temperatures of the gas exceed 
significantly $T_{\rm CMB}$ already at redshifts 14 and 10, respectively.
In the following we will focus on the radio emission signal expected from our 
simulations in the limit $T_S\gg T_{\rm CMB}$. 

\subsection{`Late reionization' model}

The first simulation describes a {\it `late reionization'}
scenario in which the galaxy emission properties are computed assuming 
a Salpeter IMF, a time-dependent spectrum typical of metal-free stars,
and an escape fraction $f_{\rm esc}=5\%$ (CSW; run S5 in CFW). 
The choice of these 
parameters leads to a reionization epoch $z_{\rm ion}\approx 8$ (defined 
here as the redshift where the volume-averaged ionization fraction, $x_v$, 
reaches unity). 
We use the expressions derived in equations (\ref{tau})-(\ref{eq:dT})
together with our simulations to account for the spatial distribution of 
structures and their different ionization levels during the reionization 
process. Maps of the antenna temperature can be thus constructed at each 
redshift based on equation (\ref{eq:dT}), with ${\cal T}$ in equation
(\ref{taucosm}) derived from the simulations along each line-of-sight. 
In Figure \ref{fig3}, the resulting $\delta T_b$ maps are shown for 
different redshifts. The maps refer to the emission produced by a slice
of the simulation box of comoving thickness $L/N^{1/3}=220\,$kpc. Highly 
ionized regions (the dark areas in the map) initially occupy only
a small fraction of the volume. At 
redshift 12, when about 90\% of the IGM is still neutral, $\delta T_b$ 
reaches values $\sim 0.1$~K in mildly overdense regions with $\delta$ greater
than a few: the antenna temperature drops rapidly with cosmic time 
as reionization proceeds. By redshift of 11 or so, several bubbles have
already overlapped as the first sources of ionizing photons are highly 
clustered. Finally, by $z=8.7$ most of the volume has been photoionized 
with only a few dense \HI patches surviving.

The distribution of $\delta T_b$ values is shown in Figure 
\ref{fig4}. At $z=18.5$, most of the IGM is still neutral, and 
the brightness temperatures reflect density 
inhomogeneities still in the linear regime. As reionization gets 
under way, the distribution becomes double peaked due to the two-phase
structure of the IGM: the rightmost peak corresponds to the neutral phase,
while the left peak is associated with ionized gas. 
The pixel fraction corresponding to $\delta T_b\ll 10^{-8}\,$K is not
drawn in the figure: this increases by two orders of magnitude as $z_{\rm ion}$
is approached. With decreasing redshift, more pixels are transferred 
from the right to the left peak. While the average 21 cm signal is two
orders of magnitude lower than that of the CMB (and will be swamped by the
much stronger non-thermal backgrounds that dominate the radio sky at
meter wavelengths and that must be removed), its fluctuations -- induced 
now both by inhomogeneities in the gas density and in the hydrogen ionized 
fraction  -- will greatly exceed those of the CMB. Brightness temperature 
fluctuations will be present both in frequency and in angle across the sky,
and should be much easier to detect.

\begin{figurehere}
\vspace{-0.5cm}
\centerline{
\psfig{figure=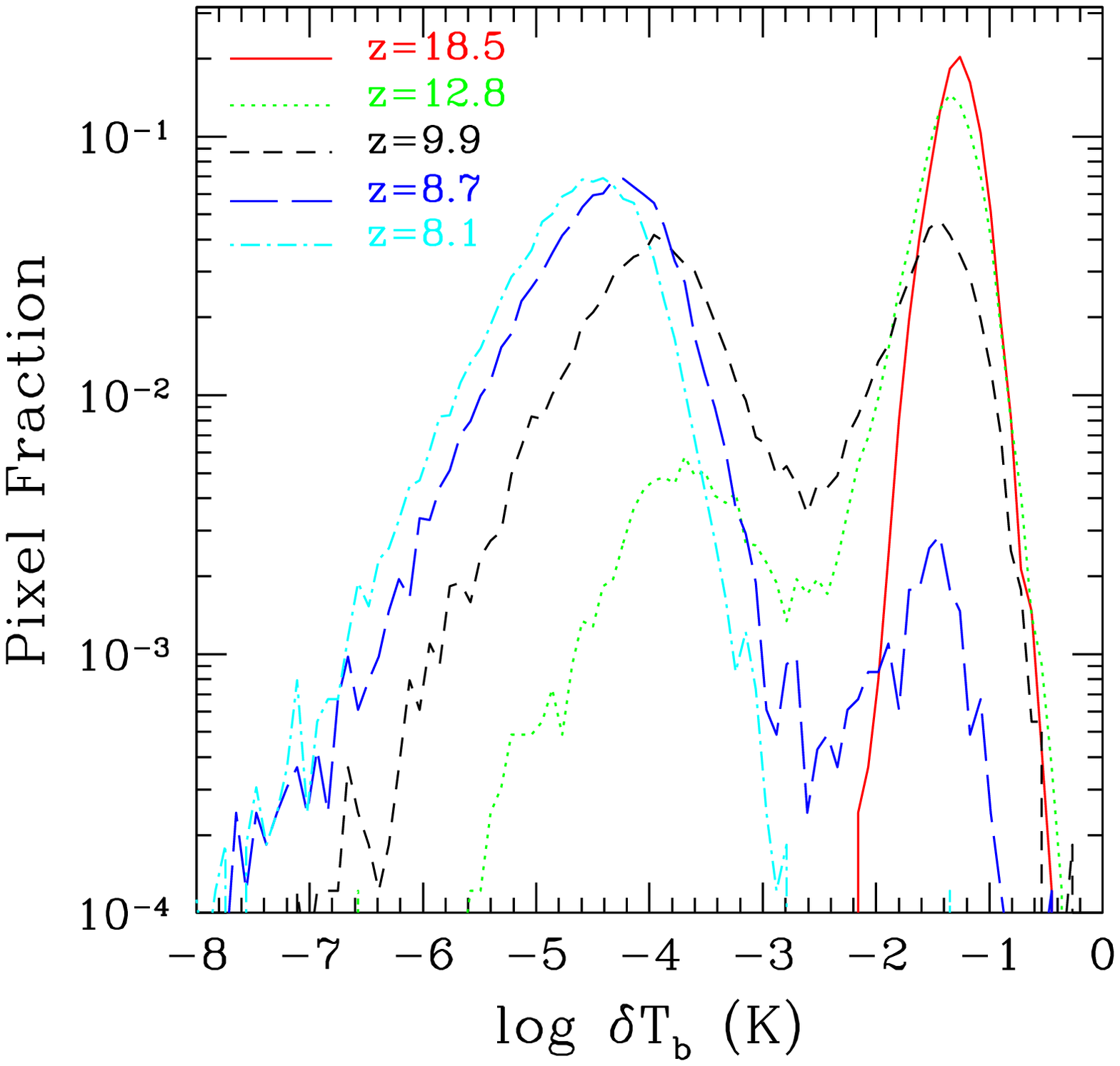,width=3.0in}}
\caption{\footnotesize Distribution in our `late reionization' simulation
box of the differential
antenna temperature $\delta T_b$ (K) at redshifts $z=$~18.5
({\it solid line}), 12.8 ({\it dotted
line}), 9.9 ({\it small-dashed line}), 8.7 ({\it long-dashed line}) and 8.1
({\it dotted-dashed line}).
}
\label{fig4}
\vspace{0.4cm}
\end{figurehere}

\begin{figurehere}
\vspace{0.3cm}
\centerline{
\psfig{figure=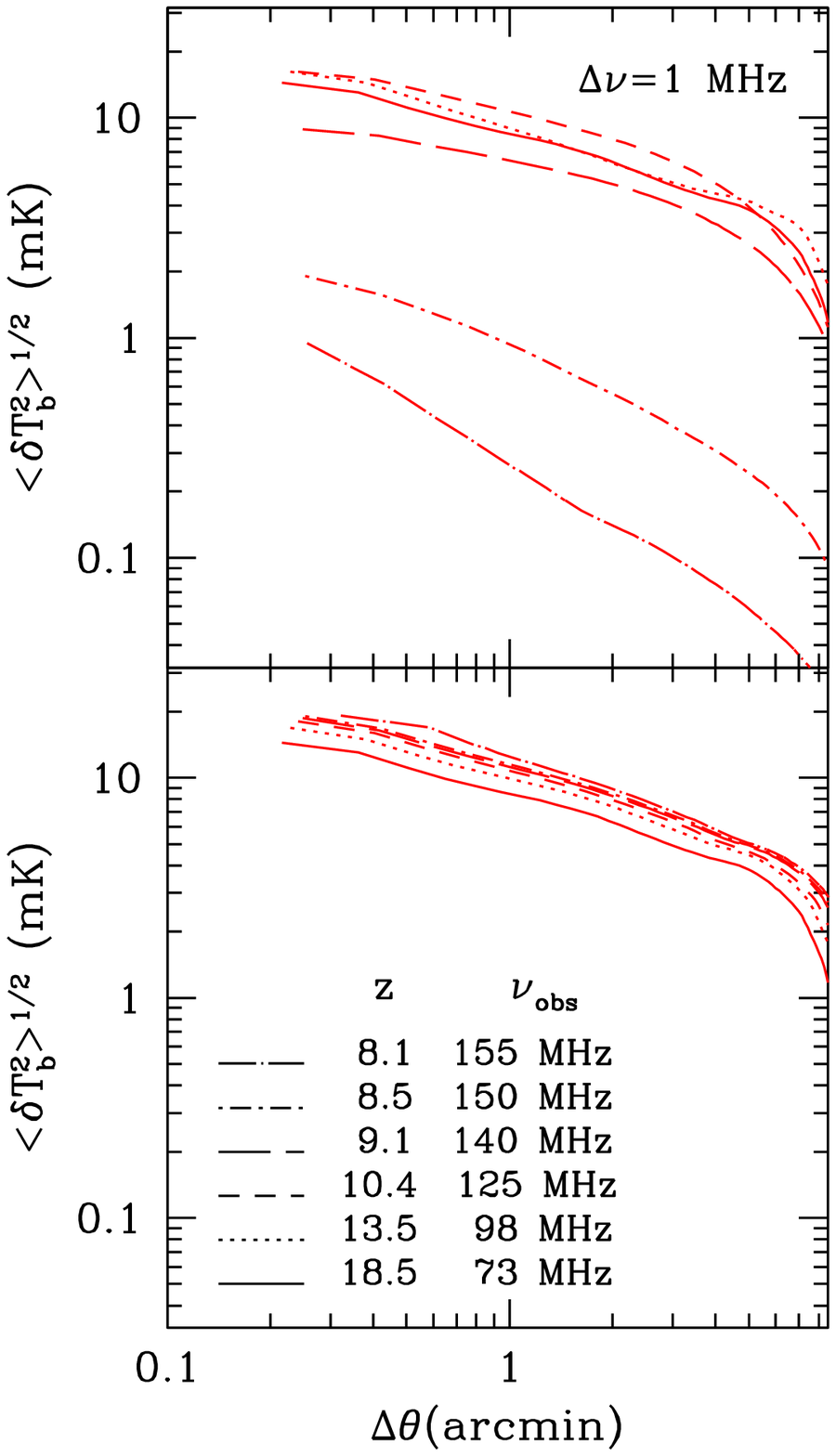,width=2.8in,height=10cm}}
\caption{\footnotesize  {\it Upper panel:} Expected rms brightness temperature
fluctuations, $\langle \delta T_b^2 \rangle^{1/2}$, in our `late reionization'
simulation box
as a function of beam size $\Delta \theta$. A fixed bandwidth $\Delta
\nu=1$~MHz has been assumed. Every curve
corresponds to a different emission redshift or, equivalently, observed
frequency $\nu_{obs}$. {\it Lower panel}: same in a test run with radiation
sources switched off. Note that the box dimension is about 10 arcmin.
}
\label{fig5}
\vspace{0.6cm}
\end{figurehere}

Figure \ref{fig5} (upper panel) shows the expected rms temperature fluctuations relative to 
the mean, $\langle \delta T_b^2 \rangle^{1/2}$, produced by a gas volume
corresponding to a given bandwidth, $\Delta \nu=\nu_{obs}\Delta z/(1+z)$, 
and angular size $\Delta \theta$, as a function of $\Delta\theta$. The 
bandwidth is fixed for each redshift and equal to $\Delta \nu=1$~MHz. 
The signal peaks at $z\sim 10.5$ ($\sim 120$ MHz), corresponding to the epoch 
when several high density neutral regions are still present, but \HII 
occupies roughly half of the volume.  As the mass variance is larger on small 
scales, at a fixed bandwidth $\langle \delta T_b^2\rangle^{1/2}$ increases 
with decreasing angular scale from about 5 to 20 mK.

\begin{figurehere}
\vspace{-0.3cm}
\centerline{
\psfig{figure=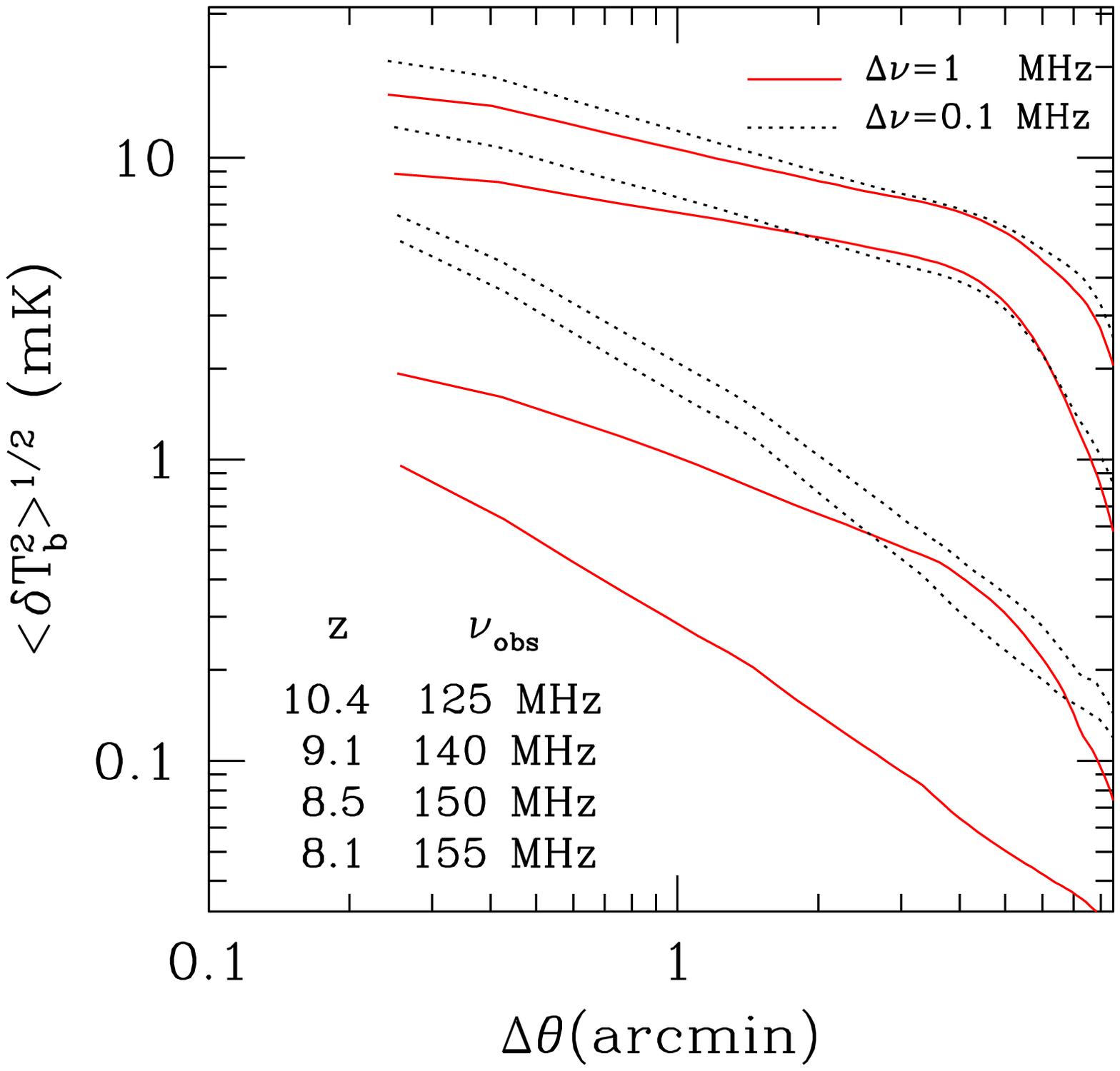,width=2.8in}}
\caption{\footnotesize Same as the upper panel of Fig. \ref{fig5} for
two different bandwidths, $\Delta\nu=0.1$ (dotted lines) and 1 MHz (solid).
}
\label{fig6}
\vspace{0.6cm}
\end{figurehere}

At lower redshift $\langle \delta T_b^2 \rangle^{1/2}$ decreases and it drops
dramatically at $z\lta 8.5$, when the IGM gets close to complete reionization.
Above redshift 11, the rms decreases again as the development of structures
in the universe is less advanced. Although inhomogeneities in the ionized 
fraction drop to zero as the IGM is mostly neutral, a patchwork of radio 
emission is produced by regions with overdensities still in the linear regime 
and by voids. Note that the errors due to the box finiteness allow a correct evaluation
of $\langle \delta T_b^2 \rangle^{1/2}$ only for $\Delta \theta \lta 8$~arcmin.
To better gauge the suppression effect of cosmological 
reionization on the sky brightness at 21 $(1+z)$ cm, we have plotted in the 
lower panel the rms temperature fluctuations
expected in the same simulation but with radiation sources switched off. 

\begin{figurehere}
\vspace{-0.3cm}
\centerline{
\psfig{figure=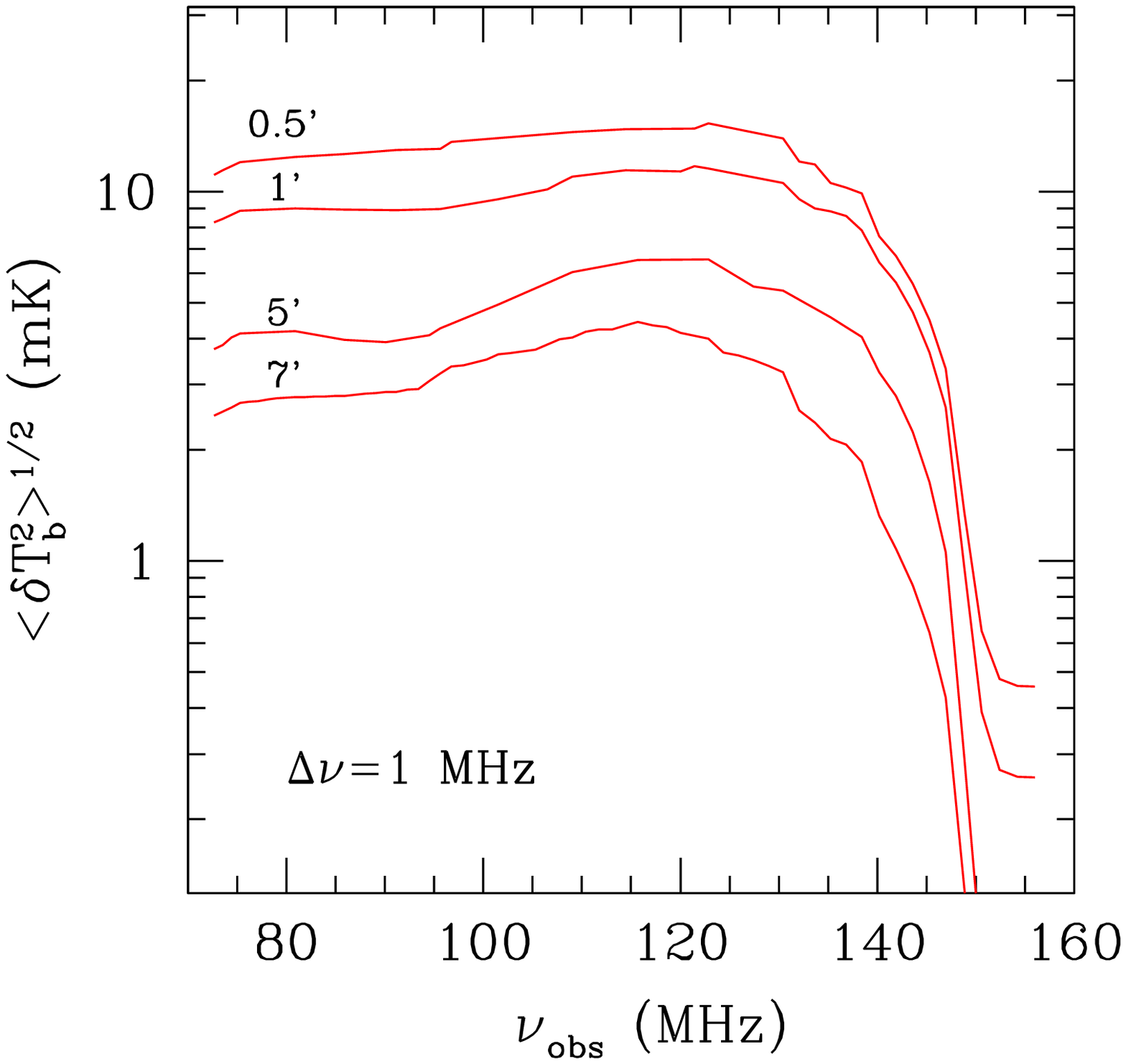,width=2.8in}}
\caption{\footnotesize Expected rms brightness temperature fluctuations
in the `late reionization' case as a function of the observed frequency,
$\nu_{obs}$, for
different choices of the beam angle, $\Delta \theta$. Notation is the same
as in Fig. \ref{fig5}. The bandwidth is $\Delta \nu=1$~MHz.
}
\label{fig7}
\vspace{0.6cm}
\end{figurehere}

We have also checked that a smaller bandwidth of $\Delta \nu=0.1$~MHz gives 
roughly the same rms at high redshifts, while this becomes larger on 
small scales as $z$ decreases (Fig. \ref{fig6}). 
This is related to the fact that inhomogeneities in the gas density and in the
hydrogen ionized fractions, which increase with decreasing redshift, are better
resolved by a smaller bandwidth.
The expected rms brightness temperature fluctuations as a 
function of the observed frequency, $\nu_{obs}$, are shown in Figure 
\ref{fig7} for our `late reionization' run, for different 
choices of beam size, and for a bandwidth $\Delta\nu=1$~MHz. The temperature
fluctuations remain approximately constant over a wide range of redshift, 
drastically declining only when the IGM becomes highly ionized.

\subsection{`Early reionization' model}

Recently, the discovery by the {\it WMAP} satellite of a large optical depth
to Thomson scattering (Kogut \etal 2003; Spergel \etal 2003) suggests that
the universe was reionized at $z_{\rm ion}>10$. We have thus repeated the
same analyses of \S\,5.1 for a simulation in which the galaxy emission
properties are computed assuming instead a Larson IMF 
(i.e. a Salpeter function with a cutoff below $5\,\msun$) and an escape 
fraction $f_{\rm esc}=20\%$ (run L20 in CFW). These parameters lead to an 
{\it `early reionization'} epoch, $z_{\rm ion}\approx 13$, more consistent with 
the {\it WMAP} observations (see CFW for an extensive discussion). 
The overall, qualitative results are the same described in the
previous Section for the `late reionization' case. Here, we limit our
discussion to the expected value of the rms brightness temperature fluctuations,
shown in Figure \ref{fig8}. The same curves in Figure \ref{fig8} and Figure
\ref{fig5} (upper panel) correspond to redshifts of equivalent volume averaged
ionization fraction, e.g., $x_v \sim 0.5$ is reached at $z=14.9$ (10.4) in the
`early (late) reionization' case and it corresponds to the short-dashed line
in both Figures.
The behavior is similar to the `late reionization' model, except that now
the neutral IGM is detectable only at radio frequencies below 100 MHz.

\begin{figurehere}
\vspace{-0.3cm}
\centerline{
\psfig{figure=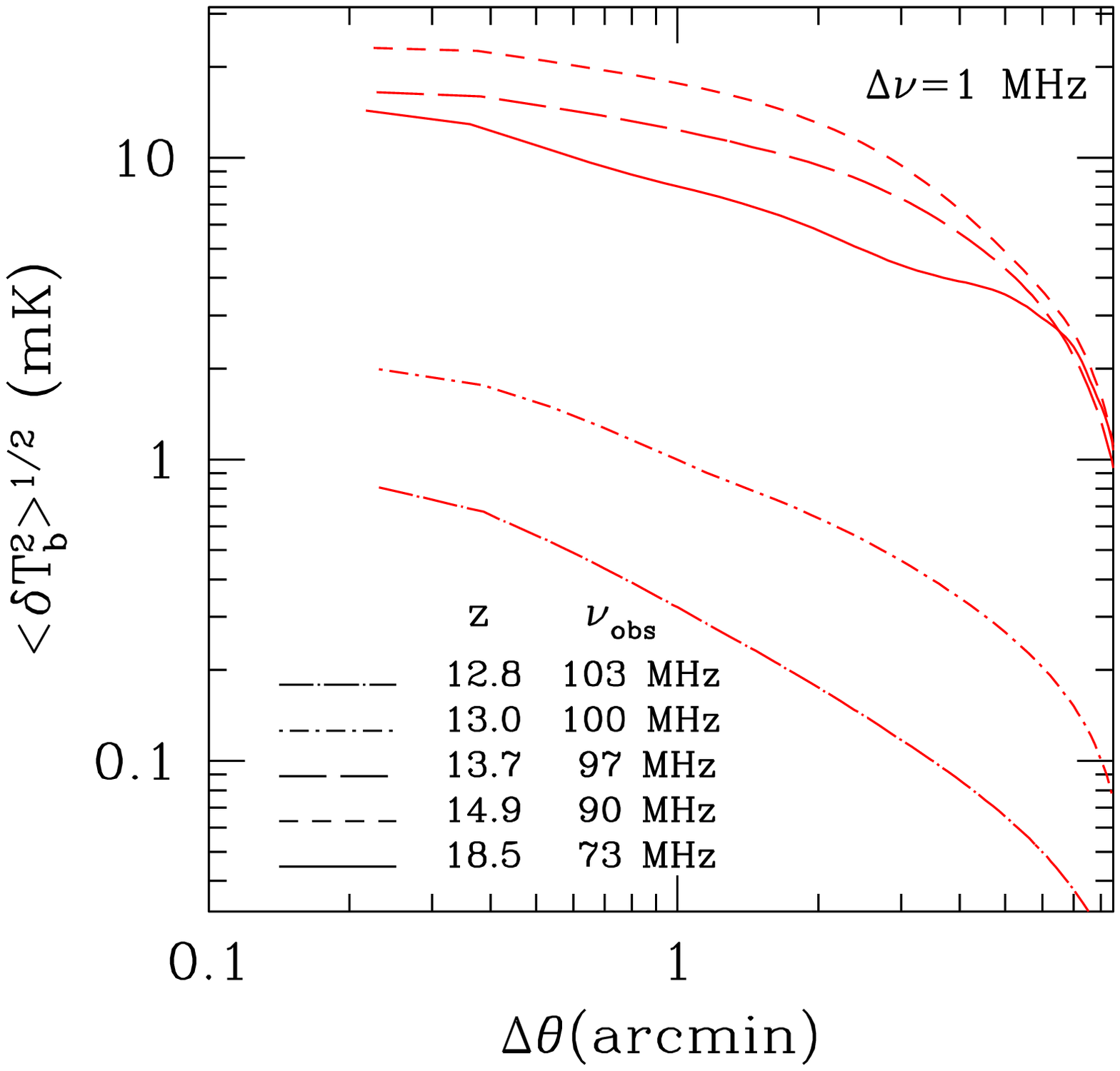,width=2.8in}}
\caption{\footnotesize 
Same as in Fig. \ref{fig5} (upper panel) but in the `early reionization' case.
Note how rms brightness temperature fluctuations on all scales now drop 
below 2 mK at $z\lta 13$, making the IGM effectively undetectable in 
21 cm radiation above radio frequencies of 100 MHz.
}
\label{fig8}
\vspace{0.6cm}
\end{figurehere}

\section{Discussion and Conclusions}

In this paper we have used numerical 
simulations of hydrogen reionization by stellar sources in the context of 
a $\Lambda$CDM cosmogony to investigate the $21\,(1+z)$~cm radio 
background expected prior to the epoch of cosmological reionization. 
Here, in addition to the emission induced by the `cosmic web' (Tozzi
\etal 2000), we can also estimate the contribution from inhomogeneities
in the hydrogen ionized fraction. A further contribution comes from
the minihalos with
virial temperatures below the threshold for atomic cooling that form in 
abundance at high redshift, and are sufficiently hot and dense to emit 
collisionally-excited 21 cm radiation (Iliev et al. 2002,2003). 
These halos are not resolved in our numerical simulations.
While the combined signal from many minihalos within a sufficiently large
volume is in principle detectable, the 21 (1+z) cm flux from these objects
will typically be dominated by the radio background from the neutral IGM
(since the \HI mass will always be larger in the non-collapsed phase; 
Oh \& Mack 2003), except in the very earliest stages of structure
formation when $J_\alpha<J_{\rm th}$, i.e. at $z>20$. At $z>12$, the predicted
brightness temperature fluctuations from minihalos on arcmin scales are 
typically much smaller than those expected from the diffuse IGM.
As reionization proceeds, feedback effects (like photoevaporation) on 
minihalos may also decrease their expected 21 cm signal.
Here, we have focused on emission from the diffuse, low density gas.
The search at 21 cm for the epoch of first light has become one of the main
science drivers of the {\it LOw Frequency ARray} ({\it LOFAR}).
Radio 21 cm tomography may
probe the topology of reionization, map the cosmic `gray age', i.e.
the transition from the post-recombination universe to one populated with
radiation sources, and effectively open up much of the universe to a
direct study of the reheating and reionization epochs.
On scales $\sim 5$ arcmin, predicted rms brightness temperature
fluctuations can exceed 5 mK for a
1 MHz bandwidth. The expected sensitivity of {\it LOFAR}  on these scales
is about 10 mK with a 1000 h integration (for a confidence level of 5
times the noise and $\Delta \nu=1$ MHz, see
http://www.astron.nl/lofar/science/), enough to detect all fluctuations above
2 $\sigma$.  While remaining an extremely
challenging project due to foreground contamination from unresolved
extragalactic radio sources  (Di Matteo \etal 2002) and free-free emission
from the same halos that reionize the universe (Oh \& Mack 2003), the
detection and imaging of large-scale structure prior to reionization
breakthrough remains a tantalizing possibility within range of the next
generation of radio arrays.

The main results of this study can be summarized as follows.

\begin{itemize}
\item At epochs when the IGM is still mainly neutral, the simulated early
galaxy population provides enough \lyac photons to decouple $T_S$ from 
$T_{\rm CMB}$. As in the same redshift range the IGM is expected to be
`warm', the 21~cm line would be seen in emission.

\item The rms temperature fluctuations relative to the mean, 
$\langle \delta T_b^2 \rangle^{1/2}$, increase with decreasing angular
scale, as variance is larger on smaller scales. The signal peaks at an
epoch when several high density neutral regions are still present, but
\HII occupies roughly half of the volume.

\item Depending on the redshift of reionization breakthrough, broad-beam observations
at frequencies $\lta 150$ MHz (below 100 MHz for the `early reionization'
scenario) with the next generation of radio telescopes 
should reveal angular fluctuations in the sky brightness temperature in the 
range 5--20 mK (1$\sigma$) on scales $\lta 5$ arcmin. 

\end{itemize}

\acknowledgments
We have benefited from discussions with T. Di Matteo, T. Ensslin,
A. Ferrara, I. Iliev, M. Rees, and P. Tozzi. B.C. thanks F. Stoehr
for helping with the simulations. Support for this work 
was provided by NSF grant AST-0205738, by NASA grant NAG5-11513,  
and by a B. Rossi visiting fellowship at the Observatory of Arcetri 
(P.M.). B.C. acknowledges
the support of the Research and Training Network ``The Physics of the 
Intergalactic Medium'' set up by the European Community under the contract 
HPRN-CT-2000-00126. 

{}

\end{document}